\documentclass[a4paper]{article}
\usepackage{amsmath,amssymb,theorem,fullpage}
\usepackage{graphicx}
\usepackage{tikz}
\usepackage[nosort]{cite}

\numberwithin{equation}{section}

\let\frac\undefined

\allowdisplaybreaks

\def\Maketitle{{\def\newpage{}\maketitle}}

\def\eq#1{\begin{equation}#1\end{equation}}
\long\def\subeq#1{\begin{subequations}#1\end{subequations}}
\def\Align#1{\begin{align}#1\end{align}}

\def\Aligned#1{\begin{aligned}#1\end{aligned}}

\def\Multline#1{\begin{multline}#1\end{multline}}

\def\pMatrix#1{\begin{pmatrix}#1\end{pmatrix}}
\def\Cases#1{\begin{cases}#1\end{cases}}

\def\Tr{\mathop{\rm Tr}\nolimits}

\def\cS{{\cal S}}
\def\cF{{\cal F}}

\def\ve{\varepsilon}
\def\vep{\varepsilon^\vphprime}
\def\vphprime{{\vphantom{\prime}}}
\def\sh{\mathop{\rm sh}\nolimits}
\def\ch{\mathop{\rm ch}\nolimits}
\def\th{\mathop{\rm th}\nolimits}

\def\lcolon{\mathopen:}
\def\rcolon{\mathclose:}
\def\bC{{\mathbb{C}}}

\def\bZ{{\mathbb{Z}}}
\def\cO{{\cal O}}

\def\hk{{\hat k}}
\def\llangle{\mathopen{\langle\!\langle}}
\def\rrangle{\mathclose{\rangle\!\rangle}}

\def\e{{\rm e}}
\def\i{{\rm i}}
\def\d{\partial}
\def\phibar{{\bar\phi}}
\def\bg{{\bar g}}
\def\cC{{\cal C}}
\def\rhobar{{\bar\rho}}

\def\rc{r^{\rm c}}
\def\rs{r^{\rm s}}

\def\ipihalf{{\textstyle{\i\pi\over2}}}

\makeatletter

\def\section{\@startsection{section}{1}{\z@}%
                                   {-3.5ex \@plus -1ex \@minus -.2ex}%
                                   {2.3ex \@plus.2ex}%
                                   {\normalfont\normalsize\bfseries}}
\def\subsection{\@startsection{subsection}{2}{\z@}%
                                     {-3.25ex\@plus -1ex \@minus -.2ex}%
                                     {1.5ex \@plus .2ex}%
                                     {\normalfont\normalsize\bfseries}}
\def\@seccntformat#1{\csname the#1\endcsname.~~}
\long\def\@makecaption#1#2{%
  \vskip\abovecaptionskip
  \sbox\@tempboxa{\small#1. #2}%
  \ifdim \wd\@tempboxa >0.9\hsize
  {\leftskip=0.05\hsize\rightskip=0.05\hsize\relax\small
    #1. #2\par}
  \else
    \global \@minipagefalse
    \hb@xt@\hsize{\hfil\box\@tempboxa\hfil}%
  \fi
  \vskip\belowcaptionskip}
\def\Appendix{\appendix
  \def\@seccntformat##1{Appendix~\csname the##1\endcsname.~~}}

\let\over\@@over
\let\atop\@@atop
\let\above\@@above
\let\overwithdelims\@@overwithdelims
\let\atopwithdelims\@@atopwithdelims
\let\abovewithdelims\@@abovewithdelims

\makeatother

\begin{document}

\def\i{{\rm i}}
\def\e{{\rm e}}

\title{Boundary form factors in the Smirnov--Fateev model\\
with a diagonal boundary $S$ matrix}

\author{Michael Lashkevich,\\[\medskipamount]
\it Landau Institute for Theoretical Physics,\\
\it 142432 Chernogolovka of Moscow Region, Russia}

\date{}

\Maketitle

\begin{abstract}
The boundary conditions with diagonal boundary $S$ matrix and the boundary form factors for the Smirnov--Fateev model on a half line has been considered in the framework of the free field representation. In contrast to the case of the sine-Gordon model, in this case the free field representation is shown to impose severe restrictions on the boundary $S$ matrix, so that a finite number of solutions is only consistent with the free field realization.
\end{abstract}

\section{Introduction}

The form factors in quantum field theory provide a natural framework for calculation of large distance asymptotics of correlation functions. In the integrable quantum field theory the form factors can be, in principle, found exactly as solution to a system of linear functional equations, called form factor axioms, as soon as the spectrum and the $S$ matrix of the model has been found~\cite{KW78,Smirnov84,Smirnovbook}. This construction admits a generalization to the case of an integrable model with a boundary. In this case the correlation functions are expressed in terms of both bulk and boundary form factors. The boundary form factors can be also found exactly~\cite{Jimbo:1994gm,Hou:1996bp}.

Here we consider a two-parametric family of integrable models proposed by Smirnov~\cite{Smirnov93} with four charged particles $z_{\ve\ve'}$ ($\ve=\pm$, $\ve'=\pm$). It means that the space of internal states of a particle is
\eq{
V=\bC^2\otimes\bC^2.
\label{Vspace}
}
The $S$ matrix of the model is factorizable, and the two-particle $S$ matrix $S_{p_1p_2}(\theta)$ is given by
\eq{
S_{p_1p_2}(\theta)=-S_{p_1}(\theta)\otimes S_{p_2}(\theta),
\label{sfSmatrix}
}
where each tensor component acts on the tensor square of the corresponding tensor component $\bC^2$ of the space~$V$. The matrix $S_p(\theta)$ is the two-soliton $S$ matrix of the sine-Gordon model with the coupling constant $\beta^2_{\rm SG}=8\pi{p\over p+1}$~\cite{ZamZam79}:
\eq{
\Aligned{
{}\span
S_p(\theta)^{++}_{++}
=-\e^{\i\delta_p(\theta)},
\qquad
S_p(\theta)^{+-}_{+-}
=-\e^{\i\delta_p(\theta)}
{\sh{\theta\over p}\over\sh{\i\pi-\theta\over p}},
\qquad
S_p(\theta)^{-+}_{+-}
=-\e^{\i\delta_p(\theta)}
{\i\sin{\pi\over p}\over\sh{\i\pi-\theta\over p}},
\\
\span
S(\theta)^{-\ve'_1,-\ve'_2}_{-\ve_1,-\ve_2}
=S(\theta)^{\ve'_1\ve'_2}_{\ve_1\ve_2},
\qquad
\delta_p(\theta)
=2\int^\infty_0{dt\over t}\,
{\sh{\pi t\over2}\sh{\pi(p-1)t\over2}\sin\theta t
\over\sh\pi t\sh{\pi pt\over2}}.
}\label{SGSmatrix}
}

The Lagrangian description of this model was found by Fateev~\cite{Fateev96}. Consider three scalar fields $\varphi_i(x)$, $i=1,2,3$ with the action
\eq{
\cS=\int d^2x\left(
{(\d_\mu\varphi_1)^2+(\d_\mu\varphi_2)^2+(\d_\mu\varphi_3)^2
\over8\pi}
+{\mu\over\pi}\left(
\cos(\alpha_1\varphi_1+\alpha_2\varphi_2)\e^{\beta\varphi_3}
+\cos(\alpha_1\varphi_1-\alpha_2\varphi_2)\e^{-\beta\varphi_3}
\right)
\right)
\label{sfmodel}
}
with the parameters $\alpha_1$, $\alpha_2$, $\beta$ satisfying the integrability condition
\eq{
\alpha_1^2+\alpha_2^2-\beta^2=1.
}
It is convenient to introduce the notation
\eq{
p_1=2\alpha_1^2,
\qquad
p_2=2\alpha_2^2,
\qquad
p_3=-2\beta^2,
\qquad
p_1+p_2+p_3=2
\label{pidef}
}
and
\eq{
\alpha_3=-\i\beta.
\label{alpha3def}
}
There are three different regimes in the theory
\subeq{\Align{
p_1,p_2>0,\quad p_3<0
&\qquad(\text{\bfseries Regime I});
\label{regionI}
\\
p_1,p_2,p_3>0
&\qquad(\text{\bfseries Regime II});
\label{regionII}
\\
p_1,p_2<0,
\quad
p_3>0
&\qquad(\text{\bfseries Regime III}).
\label{regionIII}
}}
The only unitary regime is the regime~I. In this regime the action possesses two topological charges
$$
Q_i={\alpha_i\over\pi}\int dx^1\,\d_0\varphi_i,
\qquad i=1,2.
$$
The values of the topological charges satisfy the conditions
$$
Q_1,Q_2\in\bZ,\qquad Q_1+Q_2\in2\bZ.
$$
In this regime the model (\ref{sfmodel}) can be identified with the scattering theory~(\ref{sfSmatrix}). Namely, the elementary particles $z_{\ve\ve'}$ correspond to the kinks with topological charges $Q_1=\ve$, $Q_2=\ve'$. The parameters $p_1$, $p_2$ from (\ref{pidef}) are those of the $S$ matrix~(\ref{sfSmatrix}). That is why we shall call the model with the action (\ref{sfmodel}) the Smirnov--Fateev (SF) model. The bulk form factors of a family of exponential fields in this model $\e^{\i a_1\varphi_1(x)+\i a_2\varphi_2(x)+b\varphi_3(x)}$ were obtained in~\cite{Fateev:2004un}.

Though the regime~II is nonunitary, it plays an important role in the free field representation. The reason is that the symmetry of the model with respect to the permutations $(\alpha_i,\varphi_i)\leftrightarrow(\alpha_j,\varphi_j)$ becomes apparent in this regime. There are three topological charges $Q_1$, $Q_2$, $Q_3$ (with the evident definition of~$Q_3$) in the regime~II. A part of this symmetry, the symmetry with respect to cyclic permutations plays an important role in the free field realization. Hence, we shall think of the subscripts $i$ of $\alpha_i$, $\varphi_i$, $Q_i$ etc.\ to belong the cyclic group~$\bZ_3$. Due to this symmetry, there are three types of charged particles $z^i_{\ve\ve'}$ with the topological charges $Q_i=0$, $Q_{i+1}=\ve$, $Q_{i-1}=\ve'$. There is also a set of bound states. In the regime~I only one of these three families $z_{\ve\ve'}=z^3_{\ve\ve'}$ survive.

Here we consider the SF model with a boundary. From the bootstrap point of view we only need a solution $R(\theta):V\to V$ to the boundary Yang--Baxter equation with the $S$ matrix~(\ref{sfSmatrix}):
$$
R_2(\theta_2)S_{12}(\theta_1+\theta_2)R_1(\theta_1)S_{21}(\theta_1-\theta_2)
=S_{12}(\theta_1-\theta_2)R_1(\theta_1)S_{21}(\theta_1+\theta_2)R_2(\theta_2),
$$
where $R_i(\theta)$ acts on the space of internal states $V_i$ of the $i$th particle, and $S_{ij}(\theta)$ acts on the tensor product $V_i\otimes V_j$. We shall restrict ourselves by the particular case of diagonal boundary $S$ matrices. In fact, we shall see that the free field representation provides solutions for the boundary $S$ matrices of very special form. First, all of them has the tensor product form
\eq{
R(\theta)=\rho(\theta)
\pMatrix{1&\\&r_1(\theta)}\otimes\pMatrix{1&\\&r_2(\theta)},
\label{Rform}
}
with some functions $\rho(\theta)$, $r_1(\theta)$, $r_2(\theta)$. Second, we have a finite number of solutions for these three functions that admit free field representation for the form factors. We see that the situation differs from that of the sine-Gordon model, where the whole one-parametric family of diagonal solutions to the boundary Yang--Baxter equations admit the free field representation~\cite{Hou:1996bp}. We shall compare the free field realizations of the SF model and of the sine-Gordon model in more detail later.

\section{Free field representation for bulk asymptotic states}

Let us recall the free field representation of the SF model~\cite{Fateev:2004un}. Consider three families of bosonic operators $a_i(t)$ ($i\in\bZ_3$), which depend on the real parameter $t$ and satisfy the commutation relations:
\eq{
[a_i(t),a_j(t')]
=t{\sh^2{\pi t\over2}\over\sh\pi t\sh{\pi p_it\over2}}
\delta(t+t')\delta_{ij}.
\label{acommut}
}
Let
\subeq{\label{allphidef}
\Align{
\phi_i(\theta;v)
&=\int^\infty_{-\infty}{dt\over\i t}\,
a_i(t)\e^{\i\theta t+\pi v|t|/4},
\label{phidef}
\\*
\phibar_i(\theta;v)
&=\int^\infty_{-\infty}{dt\over\i t}\,{\sh\pi t\over\sh{\pi t\over2}}
a_i(t)\e^{\i\theta t+\pi v|t|/4},
\label{phibardef}
\\
\phi^{(\pm)}_i(\theta;v)
&=2\int^\infty_0{dt\over\i t}\,
\sh{\pi p_it\over2}
a_i(\pm t)\e^{\pm\i\theta t+\pi vt/4}
\label{phipmdef}
}
and
\eq{
\chi^{(\pm)}_i(\theta)
=\phi^{(\pm)}_i(\theta;2-p_i)
+\phi^{(\pm)}_{i+1}(\theta;p_{i+1}-2)
-\phi^{(\pm)}_{i+2}(\theta;p_{i+1}-p_i).
\label{chidef}
}}
Let us also introduce three central elements (`zero modes') $\hk_i$, $i=1,2,3$. Define the vacuum $|0\rangle_{k_1,k_2,k_3}$:
\eq{
a_i(t)|0\rangle_{k_1,k_2,k_3}=0
\text{~~for $t\ge0$},
\qquad
\hk_i|0\rangle_{k_1,k_2,k_3}=k_i|0\rangle_{k_1,k_2,k_3}.
\label{vacdef}
}
The Fock space $\cF_{k_1,k_2,k_3}$ is defined as the space spanned by the vectors
$$
a_{i_1}(-t_1)\ldots a_{i_n}(-t_n)|0\rangle_{k_1,k_2,k_3},
\qquad
t_1,\ldots,t_n>0,
\qquad
n=0,1,2,\ldots.
$$
The definition of normal ordering $\lcolon\ldots\rcolon$ is evident. The conjugate vacuum ${}_{k_1,k_2,k_3}\langle0|$ is defined as
\eq{
{}_{k_1,k_2,k_3}\langle0|a_i(-t)=0
\text{~~for $t\ge0$},
\qquad
{}_{k_1,k_2,k_3}\langle0|\hk_i={}_{k_1,k_2,k_3}\langle0|k_i,
\qquad
{}_{k_1,k_2,k_3}\langle0|0\rangle_{k_1,k_2,k_3}=1.
}
The `bare' vertex operators are defined as
\subeq{\label{VIdef}\Align{
V_i(\theta)
&=\lcolon\exp\left(
\i\phi_{i+1}(\theta;p_{i+1})+\i\phi_{i+2}(\theta;-p_{i+2})
\right)\rcolon,
\label{Vdef}
\\*
I^{(\pm)}_i(\theta)
&=\lcolon\exp(-\i\phibar_i(\theta;p_i)
\pm\i\chi^{(\pm)}_i(\theta))\rcolon.
\label{Idef}
}}

These operators satisfy the following relations:
\subeq{\label{opprods}\Align{
V_i(\theta')V_j(\theta)
&=g_{ij}(\theta-\theta')\lcolon V_i(\theta)V_j(\theta')\rcolon,
\label{VVprod}
\\*
V_i(\theta')I^{(\pm)}_j(\theta)
&=w^{(\pm)}_{ij}(\theta-\theta')\lcolon V_i(\theta')I^{(\pm)}_j(\theta)\rcolon,
\label{VIprod}
\\
I^{(\pm)}_j(\theta')V_i(\theta)
&=w^{(\mp)}_{ij}(\theta-\theta')\lcolon V_i(\theta)I^{(\pm)}_j(\theta')\rcolon,
\label{IVprod}
\\*
I^{(A)}_i(\theta')I^{(B)}_j(\theta)
&=\bg^{(AB)}_{ij}(\theta-\theta')
\lcolon I^{(A)}_i(\theta')I^{(B)}_j(\theta)\rcolon.
\label{IIprod}
}}
The functions $g_{ij}$, $w_{ij}^{(\pm)}$, $\bg_{ij}^{AB}$ can be found in the Appendix~\ref{opprods-func}.

The screening operators read
\eq{
S_i(k,\kappa|\theta)
=c_i\int_{\cC_i}{d\gamma\over2\pi\i}\,
(I^{(+)}_i(\gamma)\e^\kappa-\i I^{(-)}_i(\gamma)\e^{-\kappa})
{\pi\e^{-k\gamma}\over\sh{\gamma-\theta-\i\pi/2\over p_i}},
\label{Ssf}
}
with some normalization constants $c_i$ (see Appendix~\ref{opprods-func}). The contour $\cC_i$ in this equation goes from $-\infty$ to $+\infty$ above the pole at the point~$\theta+\i\pi/2$. As for the poles related to other operators, the contour goes below all poles arising due to the operators standing to the left of the screening operator~$S_i$ and above the poles related to the operators standing to the right of~$S_i$. The screening operators commute,
\eq{
[S_i(k_i,\kappa_i|\theta_1),S_j(k_j,\kappa_j|\theta_2)]=0,
\label{SiSjcommut}
}
subject to the condition
\eq{
\kappa_i=-{\i\pi\over4}(p_ik_i+p_{i+1}k_{i+1}-p_{i+2}k_{i+2})
\label{kappadef}
}
for all $i\in\bZ_3$. Let
\eq{
\hat\kappa_i=-{\i\pi\over4}(p_i\hk_i+p_{i+1}\hk_{i+1}-p_{i+2}\hk_{i+2})
\label{hatkappadef}
}

We also need an auxiliary algebra generated by two elements $\rho$ and $\omega$ with the relations
\eq{
\omega^2=\rho^2=1,
\quad
\omega\rho=-\rho\omega,
\qquad
\Tr\rho=\Tr\omega=0.
\label{rhoomegaalg}
}
The corner Hamiltonian and the vertex operators read%
\footnote{We could remove all $\hk_i$, $\hat\kappa_i$ from these formulas by redefining the fields: $-\i{\hk_i\over2}\theta+\phi_i(\theta;v)\to\phi_i(\theta;v)$, $\i\hk_i\theta+\phibar_i(\theta;v)\to\phibar_i(\theta;v)$, $-{\pi\hk_i\over4}+\phi^{(\pm)}(\theta;v)\to\phi^{(\pm)}(\theta;v)$, but it is instructive and convenient to write them separately from the fields.}
\subeq{\label{Zisf}\Align{
H&=\int^\infty_0dt\,\sum^3_{i=1}
{\sh\pi t\sh{\pi p_it\over2}
\over\sh^2{\pi t\over2}}a_i(-t)a_i(t),
\label{Hsf}
\\
Z^i_{++}(\theta)
&=\omega V_i(\theta)\e^{(\hk_{i+1}+\hk_{i+2})\theta/2},
\label{Zippsf}
\\
Z^i_{-+}(\theta)
&=\omega\rho V_i(\theta)
S_{i+1}(\hk_{i+1},\hat\kappa_{i+1}|\theta)
\e^{(\hk_{i+1}+\hk_{i+2})\theta/2},
\label{Zimpsf}
\\
Z^i_{+-}(\theta)
&=-\omega\rho V_i(\theta)
S_{i+2}(\hk_{i+2},\hat\kappa_{i+2}|\theta-{\textstyle{\i\pi p_{i+2}\over2}})
\e^{(\hk_{i+1}+\hk_{i+2})\theta/2},
\label{Zipmsf}
\\
Z^i_{--}(\theta)
&=-\omega V_i(\theta)
S_{i+1}(\hk_{i+1},\hat\kappa_{i+1}|\theta)
S_{i+2}(\hk_{i+2},\hat\kappa_{i+2}|\theta-{\textstyle{\i\pi p_{i+2}\over{i+2}}})
\e^{(\hk_{i+1}+\hk_{i+2})\theta/2}.
\label{Zimmsf}
}}
These operators satisfy the algebra
\subeq{\label{HZalgebraII}\Align{
[H,Z^i_{\vep\ve'}(\theta)]
&=\i{d\over d\theta}Z^i_{\vep\ve'}(\theta)
-\i\Omega^i_{\vep\ve'}Z^i_{\vep\ve'}(\theta),
\qquad
\Omega^i_{\vep\ve'}={\ve k_{i+1}+\ve'k_{i+2}\over2},
\label{HZcommut}
\\
Z^i_{\vep_1\ve'_1}(\theta_1)Z^i_{\vep_2\ve'_2}(\theta_2)
&=-\sum_{\vep_3\ve'_3\vep_4\ve'_4}
S_{p_{i+1}}(\theta_1-\theta_2)^{\ve_3\ve_4}_{\ve_1\ve_2}
S_{p_{i+2}}(\theta_1-\theta_2)^{\ve'_3\ve'_4}_{\ve'_1\ve'_2}
Z^i_{\vep_4\ve'_4}(\theta_2)Z^i_{\vep_3\ve'_3}(\theta_1),
\label{Ziicommut}
\\
Z^i_{\ve\ve'_1}(\theta_1)Z^{i+1}_{\ve'_2\ve''}(\theta_2)
&=\ve\ve''\sum_{\ve'_3\ve'_4}
\hat S_{p_{i+2}}(\theta_1-\theta_2)^{\ve'_3\ve'_4}_{\ve'_1\ve'_2}
Z^{i+1}_{\ve'_4\ve''}(\theta_2)Z^i_{\ve\ve'_3}(\theta_1).
\label{Ziip1commut}
}}
Here $\hat S_p(\theta)=\i\th\left({\theta\over2}+{\i\pi p\over4}\right)S_p\left(\theta+{\i\pi p\over2}\right)$.

We see that the commutation relation for the operators $Z_{\ve\ve'}(\theta)\equiv Z^3_{\ve\ve'}(\theta)$ contains just the $S$ matrix~(\ref{sfSmatrix}). The operators $Z_{\ve\ve'}(\theta)$ describe the elementary particles in the model in the unitary regime~I, while the whole set of operator $Z^i_{\ve\ve'}(\theta)$, $i\in\bZ_3$, describes the set of elementary particles in the `symmetric' regime~II.

As it was clarified in~\cite{Lukyanov:1993hc,JMbook}, the products of vertex operators $Z^{i_1}_{\vep_1\ve'_1}(\theta_1)\ldots Z^{i_N}_{\vep_N\ve'_N}(\theta_N)$, being operators in the angular quantization scheme, are in the one-to-one correspondence with the $N$-particle eigenstates of the Hamiltonian of the system. The bulk form factors are given in terms of traces of such products.

Introduce a notation
\eq{
\llangle X\rrangle_{k_1,k_2,k_3}
={\Tr_{\cF_{k_1,k_2,k_3}}(\e^{-2\pi H}X)
\over\Tr_{\cF_{k_1,k_2,k_3}}(\e^{-2\pi H})}.
\label{doubleangles}
}
For short, we shall use the notations $I=(i,\ve,\ve')$, $I_n=(i_n,\ve_n,\ve'_n)$ etc. Besides, let $\bar I=(i,-\ve,-\ve')$, i.~e.\ the particle $z_{\bar I}$ is the antiparticle to the particle~$z_I$. Then the function
\eq{
F_{k_1k_2k_3}(\theta_1,\ldots,\theta_N)_{I_1\ldots I_N}
=\llangle Z_{I_N}(\theta_N)\ldots Z_{I_1}(\theta_1)\rrangle_{k_1,k_2,k_3}
\label{Fk1k2k3}
}
satisfy the form factor axioms and we have
\Multline{
\langle\theta'_1J_1,\ldots,\theta'_{N'}J_{N'}|
\e^{\i\sum^3_{i=1}k_i\alpha_i\varphi_i(0)}
|\theta_1I_1,\ldots,\theta_NI_N\rangle
\\
=\e^{\i{\pi\over2}\omega}N_{k_1k_2k_3}
F_{k_1k_2k_3}(\theta_1-\ipihalf,\ldots,\theta_N-\ipihalf,
\theta'_{N'}+\ipihalf,\ldots,\theta'_1+\ipihalf)
_{I_1\ldots I_N\bar J_{N'}\ldots\bar J_1}
\label{bulkff}
}
for $\theta'_1<\ldots<\theta'_{N'}$,
$\theta_1<\ldots<\theta_N$. Here
\eq{
\omega\sum^N_{n=1}\Omega_{I_n}-\sum^{N'}_{n=1}\Omega_{I_n}.
\label{omegadef}
}
The function $N_{k_1k_2k_3}$ is the normalization factor found by Baseilhac and Fateev~\cite{Baseilhac:1998eq}.

\section{Systems with boundary. General description}

Now let us discuss the notion of boundary form factors. A model with a boundary can be formulated in two ways: with a time-like boundary condition and with a space-like boundary condition~\cite{Ghoshal:1993tm}. Consider first a model with a time-like boundary (Fig.~\ref{time-like}).
\begin{figure}[t]
\hbox to\textwidth{\hfill
\parbox{.4\textwidth}{$$
\begin{tikzpicture}[>=stealth]
\fill[color=lightgray] (0,-2) rectangle (1.5,2);
\draw[very thick,->] (0,-2) -- (0,2)
  node [anchor=north,xshift=8pt] {$t$};
\draw (0,-1) node[right] {$b_1$};
\draw (0,1) node[right] {$b_2$};
\draw (-3,0) -- (.4,0);
  \draw[very thick,->] (0.4,0) -- (0.5,0) node [above,xshift=4pt] {$x$};
\filldraw (0,0) circle (2.5pt);
\draw (.8,-.5) node {$\cO\,\rlap{$(0)$}$};
\node (orig) at (0,0) [circle] {};
\node (O) at (.8,-.5) [circle] {$\phantom{\cdot}$};
\draw [->,red] (O) -- (orig);
\begin{scope}[blue,->]
\node[circle] (bigorig) at (0,0) {$\phantom{\tikz\draw (0,0) circle (.5cm);}$};
\draw (-3,-2) -- (bigorig);
\draw (-2.2,-2) -- (bigorig);
\draw[-] (-0.9,-1.5) node {$\cdots$};
\draw (-.4,-2) -- (bigorig);
\draw (bigorig) -- (-2.4,2);
\draw[-] (-1,1.5) node {$\cdots$}; 
\draw (bigorig) -- (-.4,2);
\end{scope}
\draw (-1.5,-2) node[below] {$|\theta_1I_1\ldots\theta_NI_N\rangle_{b_1}$};
\draw (-1.5,2) node[above]
  {$|\theta'_1J_1\ldots\theta'_{N'}J_{N'}\rangle_{b_2}$};
\end{tikzpicture}
$$
\caption{The time-like boundary.\hfill\break
$b_1$, $b_2$ are boundary conditions.}
\label{time-like}
}%
\hfill\hfill
\parbox{.4\textwidth}{
$$
\begin{tikzpicture}[>=stealth]
\fill[lightgray] (-2,0) rectangle (2,1.3);
\draw[very thick,->] (-2,0) -- (2,0) node[above,xshift=-5pt,yshift=2pt] {$x$};
\draw (0,-3) -- (0,.4);
\draw[very thick,->] (0,.4) -- (0,.5) node[right,xshift=2pt,yshift=-3pt] {$t$};
\fill (0,0) circle (2.5pt);
\node (orig) at (0,0) [circle] {};
\node (O) at (-.7,.5) [circle] {$\phantom{\cdot}$};
\draw[red,->] (O) node[black] {$\llap{$\cO_{b_2b_1}$}(0)$} -- (orig);
\draw (-1.9,1) node[right] {$|B\rangle$: boundary state};
\draw (0,-3.1) node[below] {$|\theta_1I_1\ldots\theta_NI_N\rangle$};
\begin{scope}[blue,->]
\node (bigorig) at (0,0) [circle]
  {$\phantom{\tikz\draw (0,0) circle (.5cm);}$};
\draw (-2,-3) -- (bigorig);
\draw (-1,-3) -- (bigorig);
\draw[-] (.6,-2) node {$\cdots\cdots$};
\draw (2.6,-3) -- (bigorig);
\end{scope}
\end{tikzpicture}
$$
\caption{The space-like boundary.}
\label{space-like}
}%
\hfill}
\end{figure}%
In this picture we consider the evolution of the system on the half line, and the boundary form factors are matrix elements of a boundary operator $\cO(t)$ at $t=0$ in the basis of eigenstates of the half line Hamiltonian:
\eq{
F_\cO\left(\theta'_1,\ldots,\theta'_{N'}\atop
\theta_1,\ldots,\theta_N\right)^{b_2,J_{1^{\vphantom{\prime}}}\ldots J_{N'}}
_{b_1,I_{1^{\vphantom{\prime}}}\ldots I_N}
={}_{b_2}\!\langle\theta'_1J_1\ldots\theta'_{N'}J_{N'}|\cO(0)
|\theta_1I_1\ldots\theta_NI_N\rangle_{b_1}
\label{tlF}
}
for $\theta'_1<\ldots<\theta'_{N'}$, $\theta_1<\ldots<\theta_N$. The boundary conditions below and above the point $t=0$, denoted as $b_1$, $b_2$ may be different, and the set of admissible operators $\cO$ depends on them.

Now consider a model with a space-like boundary (Fig.~\ref{space-like}). From the point of view of the functional integral, this picture differs from the first one just by a rotation in the corresponding Euclidean space. Nevertheless, the Hamiltonian description is quite different. We have to consider the evolution of the system on the whole line, but it inevitably ends with a special boundary state~$|B\rangle$. Hence, the form factors are matrix elements of some bulk operator $\cO_{b_2b_1}(0)$ between the eigenstates of the bulk Hamiltonian and the boundary state:
\eq{
F^B_{\cO_{b_2b_1}}(\theta_1,\ldots,\theta_N)_{I_1\ldots I_N}
=\langle B|\cO_{b_2b_1}(0)|\theta_1I_1\ldots\theta_NI_N\rangle.
\label{slF}
}
for $\theta_1<\ldots<\theta_N$. Here the boundary conditions $b_1$, $b_2$ are the result of the right action of the operator $\cO_{b_2b_1}(0)$ to the boundary bra-vector $\langle B|$.

Consider the functions in the l.~h.~s.\ of (\ref{tlF}), (\ref{slF}) as analytic functions of complex rapidities. Then these two functions are related by a rotation of the Euclidean space:
\eq{
F_\cO\left(\theta'_1,\ldots,\theta'_{N'}\atop
\theta_1,\ldots,\theta_N\right)^{b_2,J_{1^{\vphantom{\prime}}}\ldots J_{N'}}
_{b_1,I_{1^{\vphantom{\prime}}}\ldots I_N}
=\e^{\i{\pi\over2}\omega}
F^B_{\cO_{b_2b_1}}(\theta_1-\ipihalf,\ldots,\theta_N-\ipihalf,
\theta'_{N'}+\ipihalf,\ldots,\theta'_1+\ipihalf)
_{I_1\ldots I_N\>-J_{N'}\>\ldots\>-J_1},
\label{tsrel}
}
The quantity $\omega$ is defined according to (\ref{omegadef}) with $\Omega_I$ being mutual locality indeces related to the {\it bulk} operator $\cO_{b_1b_2}$.

Now let us consider the form factors in the angular quantization picture. Let us look again at Fig.~\ref{space-like}. The right half of the $x$ axis with the boundary condition $b_1$ is associated to an angular boundary ket-vector $|b_1\rangle_\cO$. The left half line is described by a bra-vector ${}_\cO\langle b_2|$. Both vectors depend on the operator~$\cO$. We have to put corner transfer matrices and vertex operators in between. Let us introduce the states
\eq{
|0_{b_1}\rangle_\cO=\e^{-{\pi\over2}H}|b_1\rangle_\cO,
\qquad
{}_\cO\langle0_{b_2}|={}_\cO\langle b_2|\e^{-{\pi\over2}H}.
\label{Bistates}
}
Do not mix these states in the angular quantization picture with the boundary state $\langle B|$ on the line. Let
\eq{
\langle X\rangle_{\cO_{b_2b_1}}
={{}_\cO\langle0_{b_2}|X|0_{b_1}\rangle_\cO
\over\sqrt{{}_\cO\langle0_{b_2}|0_{b_2}\rangle_\cO}
\>\sqrt{{}_\cO\langle0_{b_1}|0_{b_1}\rangle_\cO}}.
\label{anglesB}
}
The $F^B$ function is given by
\eq{
F^B_{\cO_{b_2b_1}}(\theta_1,\ldots,\theta_N)_{I_1\ldots I_N}
=N^B_{\cO_{b_2b_1}}
\langle Z_{I_N}(\theta_N)\ldots Z_{I_1}(\theta_1)\rangle_{\cO_{b_2b_1}}.
\label{FB-Z}
}
Here $N^B_{\cO_{b_2b_1}}$ is the normalization constant. Similarly,
\eq{
F_\cO\left(\theta'_1,\ldots,\theta'_{N'}\atop
\theta_1,\ldots,\theta_N\right)^{b_2,J_{1^{\vphantom{\prime}}}\ldots J_{N'}}
_{b_1,I_{1^{\vphantom{\prime}}}\ldots I_N}
=N^B_{\cO_{b_2b_1}}
\langle
\e^{{\pi\over2}H}Z_{\bar J_1}(\theta'_1)\ldots Z_{\bar J_{N'}}(\theta'_{N'})
\e^{-\pi H}Z_{I_N}(\theta_N)\ldots Z_{I_1}(\theta_1)\e^{{\pi\over2}H}
\rangle_{\cO_{b_2b_1}}.
\label{tlFB-Z}
}

The states $|0_b\rangle_\cO$, ${}_\cO\langle0_b|$ satisfy the relations
\eq{
\Aligned{
Z_I(\theta)|b\rangle_\cO
&=\sum_JR_b(\theta)^J_IZ_J(-\theta)|b\rangle_\cO,
\\
{}_\cO\langle b|Z_I(-\theta)
&=\sum_J{}_\cO\langle b|Z_J(\theta)R_b(\theta)^{\bar I}_{\bar J}.
}\label{vacsdef}
}
The boundary $S$ matrix $R_b(\theta)^J_I$ depends on the boundary condition~$b$. With given $Z_I(\theta)$ these equations can be used to find the bosonization of the vectors $|b\rangle_\cO$, ${}_\cO\langle b|$. Up to now, it is only known how to do it in the case of diagonal boundary $S$ matrix.%
\footnote{In the lattice theory the correlation functions and form factors of some models with nondiagonal boundary $R$ and $S$ matrices can be calculated by means of the vertex-face correspondence~\cite{Hara1999}.}

\section{Free field representation for boundary states}

Following the guidelines of~\cite{Jimbo:1994ja} let us search the state $|b\rangle_{k_1,k_2,k_3}$ in the form of a coherent state:
\eq{
|b\rangle_{k_1,k_2,k_3}=\e^F|0\rangle_{k_1,k_2,k_3},
\qquad
F=\sum^3_{i=1}\int^\infty_0{dt\over t}\,
\left(-{1\over2}K_i(t)a_i^2(-t)+\beta_i(t)a_i(-t)\right)
\label{0Bdef}
}
with some functions $K_i(t)$, $\beta_i(t)$. For shorthand, we often omit the subscript $k_1,k_2,k_3$ below. The corresponding bra-vector is defined as
\eq{
{}_{k_1,k_2,k_3}\langle b|
={}_{k_1,k_2,k_3}\langle0|\,\e^{F^*},
\label{0Bbradef}
}
where the star means the antiautomorphism
\eq{
z^*=\bar z\quad(z\in\bC),
\qquad
\hk_i^*=\hk_i,
\qquad
a_i^*(t)=-a_i(-t),
\label{stardef}
}
where bar means complex conjugate.

We expect that $Z^i_{++}(\theta)|0_b\rangle=R_i(\theta)^{++}_{++}Z^i_{++}(-\theta)|0_b\rangle$. Since $Z^i_{++}$ is an exponent of free fields, the functions $K_i(t)$ must be chosen in such a way that $a_i(t)|b\rangle=(-a_i(-t)+\ldots)|b\rangle$, where dots mean a $c$-number function of~$t$. This fixes $K_i(t)$ uniquely:
\eq{
K_i(t)={\sh\pi t\sh{\pi p_it\over2}\over\sh^2{\pi t\over2}}.
\label{Kidef}
}
With this definition we have
\eq{
\exp\left(\int^\infty_{-\infty}{dt\over t}\,f(t)a_i(t)\right)|b\rangle
=g_i[f](\theta)
\exp\left(\int^\infty_{-\infty}{dt\over t}\,f(-t)a_i(t)\right)|b\rangle,
\label{expreflection}
}
where
\Align{
\log g_i[f](\theta)
&=\int^\infty_0{dt\over t}\,
{f(t)-f(-t)\over K_i(t)}\left(\beta_i(t)
-{f(t)+f(-t)\over2}\right)
\notag
\\
&=\int^\infty_0{dt\over t}\,
\left({\beta_i(t)(f(t)-f(-t))\over K_i(t)}
-{f^2(t/2)-f^2(-t/2)\over2K_i(t/2)}\right),
\label{gifdef}
}
if $f(t)-f(-t)=O(t)$ as $t\to0$.

Roughly speaking, the reflection at the vector $|b\rangle_{k_1,k_2,k_3}$ is of the form: $\phi_i(\theta)\to\phi_i(-\theta)+\ldots$,
$\phibar_i(\theta)\to\phibar_i(-\theta)+\ldots$,
$\phi_i^{(\pm)}(\theta)\to\phi_i^{(\mp)}(\theta)+\ldots$. Hence
\subeq{\label{barereflection}
\Align{
V_i(\theta)|b\rangle_{k_1,k_2,k_3}
&=\rho_i(\theta)V_i(-\theta)|b\rangle_{k_1,k_2,k_3}
\label{Vreflection}
\\
I^{(+)}_i(\theta)|b\rangle_{k_1,k_2,k_3}
&=\rhobar_i(\theta)I^{(-)}_i(-\theta)|b\rangle_{k_1,k_2,k_3}.
\label{Ipmreflection}
}}
In particular,
\eq{
\log\rhobar_i(\theta)
=\sum_{j=0}^2\int^\infty_0{dt\over t}\,
((A_{ij}(t)\beta_{i+j}(t)+B_{ij}(t))\e^{\i\theta t}
+(C_{ij}(t)\beta_{i+j}(t)+D_{ij}(t))\e^{-\i\theta t})
\label{logridef}
}
with some functions $A_j(t),\ldots,D_j(t)$ listed in the Appendix. To get reasonable reflection of the screening operators (\ref{Ssf}) we demand
$$
(I^{(+)}_i(\theta)\,\e^{\kappa_i}-\i I^{(-)}_i(\theta)\,\e^{-\kappa_i})
|b\rangle_{k_1,k_2,k_3}
=(\rhobar(\theta)I^{(-)}_i(-\theta)\,\e^{\kappa_i}
-\i\rhobar^{-1}(-\theta)I^{(+)}_i(-\theta)\,\e^{-\kappa_i})
|b\rangle_{k_1,k_2,k_3}
$$
to coincide with
$$
\i\e^{2\kappa_i}\rhobar_i(\theta)
(I^{(+)}_i(-\theta)\,\e^{\kappa_i}-\i I^{(-)}_i(-\theta)\,\e^{-\kappa_i})
|b\rangle_{k_1,k_2,k_3}.
$$
Therefore
$$
\i\e^{3\kappa_i}\rhobar_i(\theta)=-\i\rhobar^{-1}_i(-\theta)\e^{-\kappa_i}
$$
or
$$
\rhobar_i(\theta)\rhobar_i(-\theta)=-\e^{-4\kappa_i}.
$$
This is only consistent with (\ref{logridef}), if
$$
\e^{4\kappa_i}=-1
$$
and
\eq{
\log\rhobar_i(\theta)
=2\i\int^\infty_0{dt\over t}E_i(t)\sin\theta t
\label{logriform}
}
with some function $E_i(t)$. We have a system of equations for $\beta_i(t)$:
\subeq{\label{betaiequations}
\Align{
\sum^2_{j=0}(A_{ij}(t)\beta_{i+j}(t)+B_{ij}(t))
&=E_i(t),
\label{betaiequation1}
\\
\sum^2_{j=0}(C_{ij}(t)\beta_{i+j}(t)+D_{ij}(t))
&=-E_i(t).
\label{betaiequation2}
}}
Take the sum of these two equations:
$$
\sum^2_{j=0}(A_{ij}(t)+C_{ij}(t))\beta_{i+j}(t)
+\sum^2_{j=0}(B_{ij}(t)+D_{ij}(t))=0.
$$
It is easy to check that this equation is non-degenerate and its only solution reads:
\eq{
\beta_i(\theta)=-\ch{\pi p_it\over4},
\qquad
E_i(t)=0,
\qquad
\rhobar_i(\theta)=1
\qquad
(i=1,2,3).
\label{rhobarbetasolution}
}
Hence, for the reflection functions $\rho_i(\theta)$ in (\ref{Vreflection}) we have
\eq{
\log\rho_i(\theta)={-2\i\int^\infty_0{dt\over t}\,
{\sh{\pi t\over4}\sh{3\pi t\over4}\sh{\pi(p_{i+1}+p_{i+2})t\over2}
\over\sh\pi t\sh{\pi p_{i+1}t\over2}\sh{\pi p_{i+2}t\over2}}
\sin\theta t}.
\label{rhoi}
}
Since $Z_{++}^i(\theta)\sim V_i(\theta)$, it gives the $++$ entries for the boundary $S$ matrices. To get other entries let us consider action of the screening operators on the boundary state: $S_i(k_i,\kappa_i|\theta)|b\rangle_{k_1,k_2,k_3}$. There are four cases that provide necessary reflection properties: $S_i(0,\pm{\i\pi\over4}|\theta)|b\rangle$, $S_i(\pm{1\over p_i},-{\i\pi\over4}|\theta)|b\rangle$ (we omit the subscripts $k_i$ for brevity). Consider, for example, the first case:
$$
\Aligned{
S_i(0,-{\textstyle{\i\pi\over4}}|\theta)|b\rangle
&=\int{d\gamma\over2\pi}(I^{(+)}_i(\gamma)-I^{(-)}_i(\gamma))
{\i^{-1/2}\pi c_i\over\sh{\gamma-\theta-\i\pi/2\over p_i}}|b\rangle
\\
&=-\int{d\gamma\over2\pi}(I^{(+)}_i(-\gamma)-I^{(-)}_i(-\gamma))
{\i^{-1/2}\pi c_i\over\sh{\gamma-\theta-\i\pi/2\over p_i}}|b\rangle
\\
&=\int{d\gamma\over2\pi}(I^{(+)}_i(\gamma)-I^{(-)}_i(\gamma))
{\i^{-1/2}\pi c_i\over\sh{\gamma+\theta+\i\pi/2\over p_i}}|b\rangle
}
$$
Taking a half of the sum of the expressions in the first and third line, we obtain the final expression. For all four cases it reads:
\subeq{\label{Si0B}
\Align{
S_i(0,-{\textstyle{\i\pi\over4}}|\theta)|b\rangle
&=\ch{\theta+\i\pi/2\over p_i}\int{d\gamma\over2\pi}\,
(I^{(+)}_i(\gamma)-I^{(-)}_i(\gamma))
{\i^{-{1\over2}}\pi c_i\sh{\gamma\over p_i}
\over\sh{\gamma-\theta-\i\pi/2\over p_i}\sh{\gamma+\theta+\i\pi/2\over p_i}}
|b\rangle,
\label{Si0B1}
\\
S_i(0,+{\textstyle{\i\pi\over4}}|\theta)|b\rangle
&=\sh{\theta+\i\pi/2\over p_i}\int{d\gamma\over2\pi}\,
(I^{(+)}_i(\gamma)+I^{(-)}_i(\gamma))
{\i^{+{1\over2}}\pi c_i\ch{\gamma\over p_i}
\over\sh{\gamma-\theta-\i\pi/2\over p_i}\sh{\gamma+\theta+\i\pi/2\over p_i}}
|b\rangle,
\label{Si0B2}
\\
S_i({\textstyle\pm{1\over p_i}},-{\textstyle{\i\pi\over4}}|\theta)|b\rangle
&=\e^{\mp{\theta+\i\pi/2\over p_i}}\int{d\gamma\over2\pi}\,
(I^{(+)}_i(\gamma)-I^{(-)}_i(\gamma))
{\i^{-{1\over2}}\pi c_i\sh{2\gamma\over p_i}
\over\sh{\gamma-\theta-\i\pi/2\over p_i}\sh{\gamma+\theta+\i\pi/2\over p_i}}
|b\rangle.
\label{Si0B34}
}}
Using the commutation relations (\ref{VIicommutations}) one can find the corresponding reflection equations for the products that enter $Z^i_{\ve\ve'}$. For example,
$$
\Aligned{
V_3(\theta)S_1(0,-{\textstyle{\i\pi\over4}}|\theta)|b\rangle
&=\ch{\theta+\i\pi/2\over p_i}\int{d\gamma\over2\pi}\,
V_3(\theta)(I^{(+)}_1(\gamma)-I^{(-)}_1(\gamma))
{\i^{-{1\over2}}\pi c_1\sh{\gamma\over p_1}
\over\sh{\gamma-\theta-\i\pi/2\over p_1}\sh{\gamma+\theta+\i\pi/2\over p_1}}
|b\rangle
\\
&=\ch{\theta+\i\pi/2\over p_i}\int{d\gamma\over2\pi}\,
(I^{(+)}_1(\gamma)-I^{(-)}_1(\gamma))V_3(\theta)
{\i^{-{1\over2}}\pi c_1\sh{\gamma\over p_1}
\over\sh{\gamma-\theta+\i\pi/2\over p_1}\sh{\gamma+\theta+\i\pi/2\over p_1}}
|b\rangle
\\
&=\rho_3(\theta)\ch{\theta+\i\pi/2\over p_i}\int{d\gamma\over2\pi}\,
(I^{(+)}_1(\gamma)-I^{(-)}_1(\gamma))V_3(-\theta)
{\i^{-{1\over2}}\pi c_1\sh{\gamma\over p_1}
\over\sh{\gamma-\theta+\i\pi/2\over p_1}\sh{\gamma+\theta+\i\pi/2\over p_1}}
|b\rangle
\\
&=\rho_3(\theta)\ch{\theta+\i\pi/2\over p_i}\int{d\gamma\over2\pi}\,
V_3(-\theta)(I^{(+)}_1(\gamma)-I^{(-)}_1(\gamma))
{\i^{-{1\over2}}\pi c_1\sh{\gamma\over p_1}
\over\sh{\gamma-\theta+\i\pi/2\over p_1}\sh{\gamma+\theta-\i\pi/2\over p_1}}
|b\rangle
\\
&=\rho_3(\theta)
{\ch{\i\pi/2+\theta\over p_i}\over\ch{\i\pi/2-\theta\over p_i}}
V_3(-\theta)S_1(0,-{\textstyle{\i\pi\over4}}|-\theta)|b\rangle.
}
$$
Similarly we can treat other products. As a result the relation (\ref{vacsdef}) takes the form
\eq{
Z^i_{\ve\ve'}(\theta)|0_b\rangle_{k_1,k_2,k_3}
=R_i(k_1,k_2,k_3|\theta)^{\ve\ve'}_{\ve\ve'}
Z^i_{\ve\ve'}(-\theta)|0_b\rangle_{k_1,k_2,k_3}.
\label{Zivevereflection}
}
Let
\eq{
\rc_i(\theta)
={\ch{\i\pi/2+\theta\over p_i}\over\ch{\i\pi/2-\theta\over p_i}},
\qquad
\rs_i(\theta)
={\sh{\i\pi/2+\theta\over p_i}\over\sh{\i\pi/2-\theta\over p_i}},
\qquad
r_i^\pm(\theta)
\equiv\e^{\mp2\theta/p_i}.
\label{rcspmdef}
}
Then
\eq{
R_i(k_1,k_2,k_3;\theta)=\rho_i(\theta)\e^{(k_{i+1}+k_{i+2})\theta}
\pMatrix{1&\\&r^i_{i+1}(\theta)}\otimes\pMatrix{1&\\&r^i_{i+2}(\theta)}
\label{Ridef}
}
with
\eq{
r^{i-1}_i(\theta)
=\Cases{\rc_i(\theta)
 &\text{for $k_i=0$, $\kappa_i=-{\i\pi\over4}$,}
 \\
 \rs_i(\theta)
 &\text{for $k_i=0$, $\kappa_i=+{\i\pi\over4}$,}
 \\
 r^{\pm}_i(\theta)
 &\text{for $k_i=\pm{1\over p}$, $\kappa_i=-{\i\pi\over4}$;}
}
\qquad
r^{i+1}_i(\theta)
=\Cases{\rs_i(\theta)
 &\text{for $k_i=0$, $\kappa_i=-{\i\pi\over4}$,}
 \\
 \rc_i(\theta)
 &\text{for $k_i=0$, $\kappa_i=+{\i\pi\over4}$,}
 \\
 r^{\pm}_i(\theta)
 &\text{for $k_i=\pm{1\over p}$, $\kappa_i=-{\i\pi\over4}$.}
}
\label{ridef}
}
The values of parameters $k_i$, $\kappa_i$ ($i=1,2,3$) must satisfy the relation~(\ref{kappadef}). With this restriction we obtain seven types of admissible boundary conditions, which we denote as $A_i$, $B_i$ ($i=1,2,3$),~$C$:
\eq{
\Aligned{
A_i:\
&|A_i\rangle=|b\rangle_{k_1,k_2,k_3}
\text{ with $k_i=p_i^{-1}$, $k_{i\pm1}=0$,}
\\
&\qquad
r^i_{i+1}(\theta)=\rs_{i+1}(\theta),
\quad
r^i_{i-1}(\theta)=\rs_{i-1}(\theta),
\\
&\qquad
r^{i+1}_{i-1}(\theta)=\rc_{i-1}(\theta),
\quad
r^{i+1}_i(\theta)=r^+_i(\theta),
\\
&\qquad
r^{i-1}_i(\theta)=r^+_i(\theta),
\quad
r^{i-1}_{i+1}(\theta)=\rc_{i+1}(\theta);
\\
B_i:\
&|B_i\rangle=|b\rangle_{k_1,k_2,k_3}
\text{ with $k_i=p_i^{-1}$, $k_{i\pm1}=-p_{i\pm1}^{-1}$,}
\\
&\qquad
r^{i\pm1}_i(\theta)=r^-_i(\theta),
\qquad
r^i_{i\pm1}(\theta)=r^{i\mp1}_{i\pm1}(\theta)=r^+_{i\pm1}(\theta);
\\
C:\
&|C\rangle=|b\rangle_{p_1^{-1},p_2^{-1},p_3^{-1}},
\\
&\qquad
r^i_{i\pm1}(\theta)=r_{i\pm1}^+(\theta)\quad(i=1,2,3).
}\label{AiBiCbc}
}
Generally, these boundary conditions are nonunitary, $R_b(-\theta)^J_I\ne\overline{R_b(\theta)^I_J}$, due to the factors~$r_i^\pm(\theta)$. The only exception is the boundary condition $A_3$ in the unitary regime~(\ref{regionI}). In this case the particles with $i=3$ only survive and their boundary $S$ matrix is unitary.

Consider now the bra-vectors:
\eq{
\Aligned{
A_i^*:\
&{}_{k_1,k_2,k_3}\langle A_i^*|={}_{k_1,k_2,k_3}\langle b|
\text{ with $k_i=p_i^{-1}$, $k_{i\pm1}=0$;}
\\
B_i^*:\
&{}_{k_1,k_2,k_3}\langle B_i^*|={}_{k_1,k_2,k_3}\langle b|
\text{ with $k_i=p_i^{-1}$, $k_{i\pm1}=-p_{i\pm1}^{-1}$;}
\\
C^*:\
&{}_{k_1,k_2,k_3}\langle C^*|={}_{p_1^{-1},p_2^{-1},p_3^{-1}}\langle b|.
}\label{AiBiCstarbc}
}
The boundary conditions $A_i^*$, $B_i^*$ ($i=1,2,3$), $C^*$ may differ from the boundary conditions $A_i$, $B_i$, $C$, and their boundary $S$ matrices are related with the `starless' boundary $S$ matrices as
\eq{
R_{b^*}(\theta)^J_I=\overline{R_b(-\theta)^{\bar I}_{\bar J}}.
\label{Rbstardef}
}
We conclude that
\eq{
A_i^*\ne A_i,
\quad
B_i^*=B_i,
\quad
C^*=C.
\label{AiBiCstarnonstar}
}

Since any vectors corresponding to different eigenvalues of the zero mode operators $\hk_1$, $\hk_2$, $\hk_3$ are orthogonal, the operators $\cO$ corresponding to the form factors obtained in such a way change the boundary condition according to the rule: $b_2=b_1^*$.

Now let us discuss the problem of the identification of these form factors with the particular operators in the field theory. Since the boundary condition $A_i$ is not realized in the free field representation, we consider any matrix element
\eq{
f^B_b(\theta_1,\ldots,\theta_N)_{I_1\ldots I_N}
={1\over\langle b^*|\e^{-\pi H}|b\rangle}
\langle b^*|\e^{-{\pi\over2}H}Z_{I_N}(\theta_N)\ldots Z_{I_1}(\theta_1)
\e^{-{\pi\over2}H}|b\rangle
\label{fBdef}
}
with $b=A_i,B_i,C$. Let us calculate the values $q_i$ ($i=1,2,3$) of the three topological charges~$Q_i$. Let $I_k=(i_k,\ve_k,\ve'_k)$. Then
\eq{
q_j=\sum^n_{k=1}\Cases{
0,&j=i_k,\\
\ve_k,&j=i_k-1,\\
\ve'_k,&j=i_k+1.}
\label{qidef}
}
Consider the set form factors (\ref{fBdef}) with given values $q_1$, $q_2$, $q_3$ of the topological charges. This set can be identified with the operator
\eq{
\cO_{q_1q_2q_3}(x^0)
=(N^B_{q_1q_2q_3})^{-1}\e^{\i\sum^3_{i=1}{q_i\over2\alpha_i}\tilde\varphi_i(x^0)},
\label{cOq1q2q3}
}
where $\tilde\varphi_i(x)$ is the dual fields, $\d^\mu\tilde\varphi_i(x)=\ve^{\mu\nu}\d_\nu\varphi_i(x)$, and $N^B_{q_1q_2q_3}$ is some normalization factor.

\section{Comparison with the sine-Gordon model}

Recall the free field representation for the sine-Gordon model~\cite{Lukyanov:1993pn,Lukyanov:1997bp,Hou:1996bp,Kojima:2002tc}. Let $a(t)$ be a family of bosonic operators with the commutation relations
\eq{
[a(t),a(t')]
=t{\sh{\pi t\over2}\sh{\pi(p+1)t\over2}\over\sh\pi t\sh{\pi pt\over2}}
\,\delta(t+t').
\label{SGacommut}
}
The parameter $p$ is just the parameter entering the $S$ matrix~(\ref{SGSmatrix}). Let
\Align{
\phi(\theta)
&=\int^\infty_{-\infty}{dt\over\i t}\,a(t)\e^{\i\theta t},
\label{SGphidef}
\\
\phibar(\theta)
&=\int^\infty_{-\infty}{dt\over\i t}\,
{\sh\pi t\over\sh{\pi t\over2}}a(t)\e^{\i\theta t}
=\phi(\theta+\i\pi/2)+\phi(\theta-\i\pi/2).
\label{SGphibardef}
}
Let $\hk$ be a `zero mode' operator. Let $|0\rangle_k$ be a bosonic vacuum:
\eq{
a(t)|0\rangle_k=0\quad(t\ge0),
\qquad
\hk|0\rangle_k=k|0\rangle_k.
\label{SGvacdef}
}
Define the `bare' vertex operator and the screening current as
\Align{
V(\theta)
&=\lcolon\e^{\i\phi(\theta)}\rcolon,
\label{SGVdef}
\\
I(\theta)
&=\lcolon\e^{-\i\phibar(\theta)}\rcolon.
\label{SGIdef}
}
The screening operator is defined as
\eq{
S(k|\theta)=c\int_{\cC}{d\gamma\over2\pi}\,
I(\gamma){\pi\e^{-k\gamma}\over\sh{\gamma-\theta-\i\pi/2\over p}}
\label{SGSdef}
}
with the contour $\cC$ going from $-\i\infty$ to $+\i\infty$ with a twist so that the point $\theta+{\i\pi\over2}$ is below it and point $\theta-{\i\pi\over2}$ is above it. The constant $c$ is given in Appendix~\ref{opprods-func}.

The corner Hamiltonian $H$ and the vertex operators $Z_\ve(\theta)$ are given by
\subeq{\label{SGHZ}
\Align{
\span
H=\int^\infty_0dt\,
{\sh\pi t\sh{\pi pt\over2}\over\sh{\pi t\over2}\sh{\pi(p+1)t\over2}}
a(-t)a(t),
\label{SGH}
\\
Z_+(\theta)
&=V(\theta)\e^{\hk\theta/2},
\label{SGZp}
\\
Z_-(\theta)
&=V(\theta)S(\hk|\theta)\e^{\hk\theta/2}.
\label{SGZm}
}}
Let
$$
\llangle X\rrangle_k
={\Tr_{\cF_k}(\e^{-2\pi H}X)\over\Tr_{\cF_k}(\e^{-2\pi H})}.
$$
Then the bulk form factors of the operator $\e^{\i a\varphi(x)}$ are given by
\eq{
F_a(\theta_1,\ldots,\theta_N)_{\ve_1\ldots\ve_N}
=N_a\llangle Z_{\ve_N}(\theta_N)\ldots Z_{\ve_1}(\theta_1)\rrangle
_{2a/\beta_{\rm SG}}.
\label{SGffbulk}
}
Here the normalization constant $N_a$ is the vacuum expectation value found in~\cite{Lukyanov:1996jj}.

Let us again search the boundary states in the form
\eq{
|b\rangle_k=\e^F|0\rangle_k,
\qquad
F=\int^\infty_0{dt\over t}\left(-{1\over2}K(t)a^2(-t)+\beta(t)a(-t)\right).
\label{SGOBdef}
}
Here
\eq{
K(t)={\sh\pi t\sh{\pi pt\over2}\over\sh{\pi t\over2}\sh{\pi(p+1)t\over2}}.
\label{SGKdef}
}
Let
\eq{
I(\theta)|b\rangle_k=\rhobar(\theta)I(-\theta)|b\rangle_k,
\qquad
\rhobar(\theta)
=(-1)^s{\ch{\theta-\i\lambda\over p}\over\ch{\theta+\i\lambda\over p}},
\qquad s\in\bZ.
\label{SGIreflection}
}
The equation for $\beta(t)$ is simple:
$$
A(t)\beta(t)+B(t)=E(t),
$$
where
\eq{
E(t)=-{\sh\lambda t\over\sh{\pi pt\over2}}+s
\label{SGEdef}
}
and
\eq{
\Aligned{
A(t)
&=-{\sh{\pi(p+1)t\over2}\over\sh{\pi pt\over2}},
\\
B(t)
&={\sh{\pi t\over2}\sh{\pi(p+1)t\over4}
\over2\sh{\pi t\over4}\sh{\pi pt\over4}}.
}\label{SGABdef}
}
Unlike the situation in the SF model, there is just one equation for one function $\beta(t)$ for, whose solution is unique for any function~$E(t)$. This is the consequence of the fact that the screening current here does not contain any nonsymmetric in the parameter $t$ fields like~$\chi^{(\pm)}_i(\theta)$ for the SF model. For $E(t)$ given by (\ref{SGEdef}) we have
\eq{
\beta(t)\equiv\beta_{s,\lambda}(t)
={\sh\lambda t-s\cdot\sh{\pi pt\over2}\over\sh{\pi(p+1)t\over2}}
+{\sh{\pi t\over2}\ch{\pi pt\over4}\over\sh{\pi t\over4}\ch{\pi(p+1)t\over4}}.
\label{SGbetasolution}
}
Denote the corresponding operator $F$ as~$F_{s,\lambda}$.

Now we should check that
\eq{
S(\theta)|b\rangle_k
=\int{d\gamma\over2\pi}\,
{\chi_{k,s,\lambda}(\theta)\psi_{k,s,\lambda}(\gamma)
\over\sh{\gamma-\theta-\i\pi/2\over p}\sh{\gamma+\theta+\i\pi/2\over p}}
I(\gamma)|b\rangle_k
\label{chipsidef}
}
with any $\gamma$-independent function $\chi_{k,s,\lambda}(\theta)$ and any $\theta$-independent function $\psi_{k,s,\lambda}(\gamma)$. The function $\rhobar(\theta)$ of the form (\ref{SGIreflection}) is consistent with this assumption in three cases:
\subeq{\label{SGXYbc}
\Align{
X_\lambda:
&\ 
|X_\lambda\rangle=\e^{F_{1,\lambda}}|0\rangle_0
\quad\lambda\text{ is arbitrary;}
\label{SGXbc}
\\
Y_\pm:
&\ |Y_\pm\rangle=\e^{F_{1,0}}|0\rangle_{\pm p^{-1}}.
\label{SGYbc}
}}
The case $k=0$, $s=0\pmod2$, $\lambda=0$, which also satisfies the condition~(\ref{chipsidef}), is equivalent to the case $k=0$, $s=1\pmod2$, $\lambda=\pm{\pi p\over2}$. Note, that the boundary conditions $Y_\pm$ are nonunitary, while the boundary condition $X_\lambda$ is unitary for real values of the parameter~$\lambda$.

Similarly, define the corresponding bra-vectors
\subeq{\label{SGXYbrabc}
\Align{
X_\lambda:
&\ \langle X_\lambda|={}_0\langle0|\,\e^{F_{1,\pi-\lambda}^*},
\quad\lambda\text{ is arbitrary;}
\label{SGXbrabc}
\\
Y_\pm:
&\ \langle Y_\pm|={}_{\pm p^{-1}}\langle0|\,\e^{F_{1,0}^*}.
\label{SGYbrabc}
}}

The reflection property of the operator $V(\theta)$ reads
\eq{
V(\theta)=\rho(\theta)V(-\theta),
\qquad
\log\rho_{s,\lambda}(\theta)=2\i\int^\infty_0{dt\over t}
\left({\beta_{s,\lambda}(t)\over K(t)}-{1\over2K(t/2)}\right)\sin\theta t.
\label{SGVreflection}
}
The boundary $S$ matrix have the form
\eq{
R_{k,s,\lambda}(\theta)
=\rho_{s,\lambda}(\theta)\e^{k\theta}
\pMatrix{1&\\
&r_{k,s,\lambda}(\theta)},
\qquad
r_{k,s,\lambda}(\theta)
={\chi_{k,s,\lambda}(\theta)\over\chi_{k,s,\lambda}(-\theta)}.
\label{SGRdef}
}
Surely, the function $r_{s,\lambda}(\theta)$ is only defined for the values of $s$ and $\lambda$ defined in~(\ref{SGXYbc}). It reads
\subeq{\label{SGXYrdef}
\Align{
X_\lambda:
&\ r_{0,1,\lambda}(\theta)
={\ch{\i(\pi/2-\lambda)+\theta\over p}
\over\ch{\i(\pi/2-\lambda)-\theta\over p}};
\label{SGXrdef}
\\
Y_\pm:
&\ r_{\pm p^{-1},1,0}(\theta)
=\e^{\mp2\theta/p}.
\label{SGYrdef}
}}
The family $X_\lambda$ corresponds to the family of Dirichlet boundary condition described in~\cite{Ghoshal:1993tm} with
$$
\beta_{SG}\varphi(t,x=0)={2\lambda-\pi\over p+1},
\qquad
|\beta_{SG}\varphi(t,x=0)|\le\pi.
$$
The boundary conditions $Y_\pm$ from the point of view of the boundary $S$ matrix correspond to the limits $\lambda\to\pm\i\infty$. Nevertheless, we want to separate them from the family $X_\lambda$ due to two reasons. First, they do not correspond to any known boundary conditions. Second, the free field representation provides finite and rather explicit expressions for boundary form factors with these boundary conditions. A peculiarity of these expression is a non-zero value of~$k$. As we have seen, in the case of the SF model such kind of boundary conditions appear inevitably.

Identification of the form factors is similar to the SF case. Consider the function
\eq{
f^B_{\cO^q_{b'b}}(\theta_1,\ldots,\theta_N)_{\ve_1\ldots\ve_N}
={\langle b'|\e^{-{\pi\over2}H}
Z_{\ve_N}(\theta_N)\ldots Z_{\ve_1}(\theta_1)
\e^{-{\pi\over2}H}|b\rangle
\over\sqrt{\langle b'|\e^{-\pi H}|b'\rangle\langle b|\e^{-\pi H}|b\rangle}},
\qquad
q=\sum^N_{n=1}\ve_n.
\label{SGfB}
}
In terms of the dual field $\tilde\varphi(x)$, $\d^\mu\tilde\varphi(x)=\ve^{\mu\nu}\d_\nu\varphi(x)$, the operator $\cO^q_{b'b}$ can be identified as
\eq{
\cO^q(x^0)=\e^{\i q{p+1\over2p}\beta_{SG}\tilde\varphi(x^0)}
\label{SGcOdef}
}
with the appropriate change of the boundary condition at the point~$x^0$.

\section{Conclusion}

A free field representation for boundary form factors of some boundary fields in the Smirnov--Fateev model with a boundary has been found. This representation is limited to the boundary conditions with a diagonal boundary $S$ matrix. It turns out that the consistency condition of the free field representation restricts the admissible boundary conditions to a finite number. This contrasts to the situation in the sine-Gordon model, where the admissible (from the point of view of the free field representation) boundary conditions form a one-parameter family. Note that this restriction is not due to the boundary Yang--Baxter equation, which only demands that
$$
r^{i+1}_i(\theta)
={\ch{\i x_i-\theta\over p_i}\over\ch{\i x_i+\theta\over p_i}},
\qquad
r^{i-1}_i(\theta)
={\sh{\i x_i-\theta\over p_i}\over\sh{\i x_i+\theta\over p_i}}
\qquad(i=1,2,3)
$$
with some values of the parameters $x_1$, $x_2$,~$x_3$. The described free field representation only admits the solutions with either the two of these parameters being equal to $-\pi/2$ and the third tending to $-\i\infty$ (the~$A_i$ and $A_i^*$ boundary conditions) or with all three tending to~$\pm\i\infty$ (the~$B_i$ and $C$ boundary conditions). It is not clear, if this restriction is physical, or it is a limitation of the free field technique. Probably, a study of consistency of higher quantum conserved currents of the model with the boundary conditions along the guidelines of~\cite{Penati:1995bs} could shed light on this problem.

Another problem to be solved is identification of the boundary $S$ matrices for the cases $A_i$, $B$ with the particular conditions in the Lagrangian form. Note, that it would be interesting to do the same for the solutions denoted above as $Y_\pm$ in the case of the sine-Gordon model. The solution to this problem could be found by studying nonlocal integrals of motion following the guidelines of~\cite{Delius:2001qh}.

\section*{Acknowledgments}

I am grateful to P.~Baseilhac for his hospitality during my stay at the University of Tours and for interesting discussions. My last visit there was supported by the program ENS--Landau. I am also grateful to T.~Miwa, M.~Jimbo and J.~Shiraishi for their hospitality at the Kyoto University and the University of Tokyo. The work was supported, in part, by the Russian Foundation of Basic Research under the grants RFBR 05--01--01007, 05--01--02934 and by the Program of Support for the Leading Scientific Schools under the grant No.~6358.2006.2.

\Appendix
\section{The functions in Eqs.~(\ref{opprods}),~(\ref{Ssf}),~(\ref{SGSdef})%
\label{opprods-func}}

The functions $g_{ij}(\theta)$ are defined as follows ($i,j$ are
understood modulo~3):
\subeq{\Align{
g_{ii}(\theta)
&=G^{-1}(p_{i+1},\theta)G^{-1}(p_{i+2},\theta),
\quad
G(p,\theta)
=\exp\int^\infty_0{dt\over t}\,{\sh^2{\pi t\over2}\ch{\pi pt\over2}
\over\sh\pi t\sh{\pi pt\over2}}\,\e^{-i\theta t},
\\*
g_{ij}(\theta)
&=G^{-1}_1(p_k,\theta)
\quad(i\ne j,~k\ne i,j),
\quad
G_1(p,\theta)
=\exp\int^\infty_0{dt\over t}\,{\sh^2{\pi t\over2}
\over\sh\pi t\sh{\pi pt\over2}}\,\e^{-\i\theta t}.
}}
Here the integrals of the form
$$
\int^\infty_0 dt\,f(t)
$$
with $f(t)$ having a pole at $t=0$ are understood as~\cite{JKM}
$$
\int_{\cC_0}{dt\over2\pi\i}\,f(t)\log(-t)
$$
with the contour $\cC_0$ going from $+\infty+\i0$ above the real axis, then
around zero, and then below the real axis to $+\infty-\i0$.

The functions $w^{(\pm)}_{ij}(\theta)$ can be expressed in terms of
the gamma-functions:
\subeq{\label{wdef}
\Align{
w^{(+)}_{ii}(\theta)
&=w^{(-)}_{ii}(\theta)=1,
\\*
w^{(+)}_{i-1,i}(\theta)
&=w(p_i,0|\theta),
\qquad
w^{(-)}_{i-1,i}(\theta)
=w(p_i,1|\theta),
\label{wpm01}
\\
w^{(+)}_{i+1,i}(\theta)
&=w^{(-)}_{i+1,i}(\theta)=w(p_i,1/2|\theta),
\label{wpm12}
}
where
\eq{
w(p,z|\theta)
=r_p^{-1}
\>{\Gamma\left({\i\theta\over\pi p}-{1\over2p}+z\right)
\over\Gamma\left({\i\theta\over\pi p}+{1\over2p}+z\right)},
\qquad r_p=\e^{(C_E+\log\pi p)/p}
\label{wpz}
}
}
with $C_E$ being the Euler constant. Note, that all these functions
have one series of poles at the points $\theta=-\i\pi+\i\pi pn$ or
$\theta=-\i\pi+\i\pi p(n+1/2)$ ($n=0,1,2,\ldots$) and one series of
zeros at the points $\theta=\i\pi-\i\pi pn$ or $\theta=\i\pi-\i\pi
p(n+1/2)$.
The functions $\bg^{(AB)}_{ij}(\theta)$ ($A,B=\pm$) read
\subeq{\label{bgdef}
\Align{
\bg^{(-+)}_{ii}(\theta)
&=\bg(p_1,0,0|\theta),
\qquad
\bg^{(--)}_{ii}(\theta)
=\bg^{(++)}_{ii}(\theta)
={\i\theta\over\pi p_i}\bg(p_i,0,1|\theta),
\qquad
\bg^{(+-)}_{ii}(\theta)
=\bg(p_i,1,1|\theta),
\label{bgiidef}
\\
\bg^{(-+)}_{i,i+1}(\theta)
&=\bg^{(+-)}_{i,i+1}(\theta)=1,
\qquad
\bg^{(--)}_{i,i+1}(\theta)
=\bg^{(++)}_{i,i+1}(-\theta)
={\theta-\i\pi(p_{i+1}-2)/2\over\theta-\i\pi p_{i+1}/2},
\qquad
\bg^{(AB)}_{i+1,i}(\theta)
=\bg^{(BA)}_{i,i+1}(-\theta)
\label{bgijprop}
}
with
\eq{
\bg(p,z_1,z_2|\theta)
=r_p^2\>
{\Gamma\left({\i\theta\over\pi p}+{1\over p}+z_1\right)
\over\Gamma\left({\i\theta\over\pi p}-{1\over p}+z_2\right)}.
\label{bgfuncdef}
}}
The constants $c_i$ are given by
\eq{
c_i=-{\e^{2(C_E+\log\pi p_i)/p_i}\over\pi^{3/2}}\,
{\Gamma(1+1/p_i)\over\Gamma(-1/p_i)}\,
G(p_i,-\i\pi).
\label{cidef}
}
The constant in the expression (\ref{SGSdef}) for the sine-Gordon model is given by
\eq{
c={\e^{\alpha_+^2(C_E+\log\pi p)}
\over\pi^2p^2}
{\Gamma\left(1+{1\over p}\right)
\over\Gamma\left(-{1\over p}\right)}
\exp\int^\infty_0{dt\over t}\,
{\sh{\pi t\over2}\sh{\pi(p+1)t\over2}
\over\sh\pi t\sh{\pi pt\over2}}\e^{-\pi t}.
\label{SGcdef}
}

We also need the commutation relations, that follow from Eqs.~(\ref{VIprod}--\ref{IIprod}), (\ref{wdef}), (\ref{bgijprop}):
\subeq{\label{VIicommutations}\Align{
V_i(\theta_1)I^{(A)}_{i+1}(\theta_2)
&={\sh{\theta_2-\theta_1-\i\pi/2\over p}
\over\sh{\theta_2-\theta_1+\i\pi/2\over p}}
I^{(A)}_{i+1}(\theta_2)V_i(\theta_1),
\label{VIii+1commut}
\\
V_i(\theta_1)I^{(A)}_{i-1}(\theta_2)
&={\ch{\theta_2-\theta_1-\i\pi/2\over p}
\over\ch{\theta_2-\theta_1+\i\pi/2\over p}}
I^{(A)}_{i-1}(\theta_2)V_i(\theta_1),
\label{VIii-1commut}
\\
I^{(A)}_i(\theta_1)I^{(B)}_{i+1}(\theta_2)
&=I^{(B)}_{i+1}(\theta_2)I^{(A)}_i(\theta_1).
\label{IIijcommut}
}}

\bigskip

\section{The functions in Eq.~(\ref{logridef})}

The functions $A_i(t),\ldots,D_i(t)$ are given by
\eq{
\Aligned{
A_{i0}(t)
&=-2K_i^{-1}(t)\,\e^{-{\pi p_it\over4}}\ch{\pi(1-p_i)t\over2},
&
B_{i0}(t)
&=-2K_i^{-1}(t/2)\,\e^{-{\pi p_it\over4}}\ch^2{\pi(1-p_i)t\over4},
\\
C_{i0}(t)
&=2K_i^{-1}(t)\,\e^{\pi p_it\over4}\ch{\pi t\over2},
&
D_{i0}(t)
&=2K_i^{-1}(t/2)\,\e^{\pi p_it\over4}\ch^2{\pi t\over4},
\\
A_{i1}(t)
&=2K_{i+1}^{-1}(t)\,\e^{\pi(p_{i+1}-2)t\over4}\sh{\pi p_{i+1}t\over2},
&
B_{i1}(t)
&=-2K_{i+1}^{-1}(t/2)\,\e^{\pi(p_{i+1}-2)t\over4}\sh^2{\pi p_{i+1}t\over4},
\\
C_{i1}(t)
&=0,
&
D_{i1}(t)
&=0,
\\
A_{i2}(t)
&=-2K_{i+2}^{-1}(t)\,\e^{\pi(p_{i+1}-p_i)t\over4}\sh{\pi p_{i+2}t\over2},
&
B_{i2}(t)
&=-2K_{i+2}^{-1}(t/2)\,\e^{\pi(p_{i+1}-p_i)t\over4}\sh^2{\pi p_{i+2}t\over4},
\\
C_{i2}(t)
&=0,
&
D_{i2}(t)
&=0.
}\label{ABCDdef}
}

\end{document}